\documentclass[11pt]{article}
\usepackage{graphicx}
\usepackage{amssymb}
\usepackage{amscd}
\usepackage{mathrsfs}
\usepackage{longtable,lscape}
\usepackage{amsthm}
\usepackage{amsfonts}
\usepackage{amsmath}
\usepackage{bbm}
\usepackage{float}
\usepackage{url}
\usepackage[title]{appendix}
\usepackage{csquotes}

\newcommand{\real}{{\mathbb R}}
\newcommand{\captionfonts}{\footnotesize}
\makeatletter  
\long\def\@makecaption#1#2{%
  \vskip\abovecaptionskip
  \sbox\@tempboxa{{\captionfonts #1: #2}}%
  \ifdim \wd\@tempboxa >\hsize
    {\captionfonts #1: #2\par}
  \else
    \hbox to\hsize{\hfil\box\@tempboxa\hfil}%
  \fi
  \vskip\belowcaptionskip}
\makeatother 
\begin{document}
\title{Why qubits are exceptional in Gleason's theorem: a Bloch-space perspective}
\author{Massimiliano Sassoli de Bianchi\vspace{0.5 cm} \\ 
        Center Leo Apostel for Interdisciplinary Studies, 
         Brussels Free University \\ 
        \normalsize\itshape
         Krijgskundestraat 33, 1160 Brussels, Belgium \\
        \normalsize
        E-Mail: \url{msassoli@vub.ac.be}
          \vspace{0.5 cm} \\ 
         \normalsize\itshape
              }
\date{}
\maketitle
\begin{abstract}
\noindent 
Gleason's theorem is often cited as establishing the Born rule from the structure of Hilbert space, yet its original proof is mathematically sophisticated and rarely accessible to physicists. In this article we present a simple route to the dimensional distinction at the heart of Gleason's result, using the generalized Bloch representation of quantum states. More precisely, we consider a restricted but instructive class of probability assignments obtained by replacing the linear dependence of the Born expression on a Bloch-space scalar product with a real function $f$. For qubits, infinitely many continuous non-Born assignments of this form satisfy normalization for every projective measurement. In dimension $N\geq3$, by contrast, the geometry of the measurement simplex imposes additional relations among the relevant scalar products. Requiring normalization for arbitrary states then leads to a Cauchy-type functional equation, which, together with boundedness, forces $f$ to be linear. 
Our argument provides a complementary geometric illustration of why the freedom present in dimension two disappears in higher dimensions, reaching this conclusion directly from the geometry of the measurement simplices and normalization, without explicitly invoking the intertwining of projectors across different orthogonal resolutions of the identity.
\end{abstract}
\medskip
{\bf Keywords}:  Gleason's theorem; Bloch representation; Born rule; Qubits

\section{Introduction}

\noindent The Born rule occupies a central place in quantum mechanics. It provides the link between the mathematical formalism of Hilbert space and experimental probabilities, assigning to a state $D$ and a projector $P$ the probability ${\rm Tr}(DP)$. In most standard presentations, however, Born's rule is introduced as an independent postulate, although this can be justified by the structure of Hilbert space itself. This is precisely the task accomplished by Gleason's theorem \cite{Gleason1957}, which shows that, for Hilbert spaces of dimension three or higher, any probability measure defined on the lattice of projection operators must necessarily be representable by a density operator via Born's rule. In other words, under very general and natural assumptions, the Hilbert space formalism forces the Born rule.

Despite its conceptual importance, Gleason's original proof is technically demanding, and the structural reason for the restriction to Hilbert-space dimension $N\geq 3$ can be obscured by the details of the proof. A useful way of formulating that reason is Gleason's notion of \emph{intertwining}: in dimensions three and higher, the same one-dimensional projector can occur in more than one orthogonal resolution of the identity, so a non-contextual probability assigned to that projector must satisfy consistency conditions across different measurement contexts. In dimension two, by contrast, a rank-one projector has a unique orthogonal complement, and this network of constraints degenerates \cite{Gleason1957}.

The purpose of this article is not to reproduce Gleason's proof, nor to derive the density-operator formalism from probability assignments. Density operators and the Bloch representation are taken as the starting point. We then ask a more limited question: if the usual Born expression is deformed by replacing its linear dependence on the Bloch-space scalar product by a function $f$, which deformations remain compatible with the normalization constraints of all projective measurements? For qubits, infinitely many non-Born choices survive. For $N\geq3$, by contrast, the geometry of the measurement simplex imposes additional relations among the scalar products entering the probability assignments. Requiring normalization for arbitrary states then produces a functional equation for $f$, whose bounded solutions are linear \cite{Cooke1985,Dvurecenskij1996,WrightWeigert2019Cauchy}.

For clarity, let us recall the standard finite-dimensional formulation of Gleason's theorem \cite{Gleason1957}. A so-called \emph{frame function} assigns a non-negative number $\mu(P)$ to each rank-one projector $P$ such that, for every orthonormal basis with projectors $\{P_i\}_{i=1}^N$, one has $\sum_i\mu(P_i)=1$. Gleason's theorem states that, for $N\geq3$, every such non-contextual assignment is of the form $\mu(P)={\rm Tr}(\rho P)$ for a unique density operator $\rho$. 

The present argument considers only a special family of candidate frame functions parameterized by $f$ and shows how, within this family, the geometry of higher-dimensional measurement simplices and the normalization requirement eliminate nonlinear probability assignments.

\section{Pauli matrices}
\label{Pauli-matrices}

\noindent We start by recalling that given an arbitrary $2\times 2$ matrix
 \begin{equation}
D=
\left[ \begin{array}{cc}
D_{11} & D_{12} \\
D_{21} & D_{22} \end{array} \right]
\end{equation}
we can always write it as a linear combination of the identity matrix 
$\mathbb{I}$ and the three Pauli spin matrices:
 \begin{equation}
\mathbb{I}=
\left[ \begin{array}{cc}
1 & 0 \\
0 & 1 \end{array} \right],\quad
\sigma_1=
\left[ \begin{array}{cc}
0 & 1 \\
1 & 0 \end{array} \right],\quad
\sigma_2=
\left[\begin{array}{cc}
0 & -i \\
i & 0 \end{array} \right],\quad 
\sigma_3=
\left[ \begin{array}{cc}
1 & 0 \\
0 & -1 \end{array} \right]
\end{equation}
Indeed, one can easily check that:
 \begin{equation}
D={1\over 2}\left[(D_{11}+D_{22})\mathbb{I}+(D_{12}+D_{21})\sigma_1+i(D_{12}-D_{21})\sigma_2+(D_{11}-D_{22})\sigma_3\right]
\label{Bloch-decomp2x2}
\end{equation}
In other words, every  $2\times 2$ matrix $D$ can be written in the form: 
\begin{equation}
D = {1\over 2}\left(r_0\mathbb{I} +\sum_{i=1}^3 r_i\sigma_i \right)= {1\over 2}\left(r_0\mathbb{I} + {\bf r}\cdot\mbox{\boldmath$\sigma$}\right)
\label{formula2X2-1}
\end{equation}
where $r_0=D_{11}+D_{22}$, $r_1=D_{12}+D_{21}$, $r_2=i(D_{12}-D_{21})$, and $r_3=D_{11}-D_{22}$. Pauli matrices have the following remarkable properties: 
\begin{equation}
{\rm det}\, \sigma_i = -1,\quad {\rm Tr} \,\sigma_i = 0,\quad {\rm Tr} (\sigma_i\sigma_k)=2\delta_{ik}
\label{pauliproperties2}
\end{equation}
Taking the trace of $D$ and $D\sigma_k$, we thus obtain 
\begin{equation}
{\rm Tr}D=r_0,\quad {\rm Tr} (D \sigma_k) =  {1\over 2}\sum_{i=1}^3 r_i {\rm Tr} (\sigma_i\sigma_k) =\sum_{i=1}^3 r_i \delta_{ik}=r_k
\label{formula2X2-2}
\end{equation}
Pauli matrices being self-adjoint, it  follows that $D^\dagger = {1\over 2}(r_0^*\mathbb{I} + {\bf r}^*\cdot\mbox{\boldmath$\sigma$})$, hence $D=D^\dagger$ requires the coefficient $r_0$ and the vector ${\bf r}$ to be real. If $D$ is also of unit trace, we have $r_0=1$, and (\ref{formula2X2-1}) becomes (we now explicitly state the dependence on the vector {\bf r}):
\begin{equation}
D({\bf r}) =  {1\over 2}\left(\mathbb{I} + {\bf r}\cdot\mbox{\boldmath$\sigma$}\right)
\label{formula2X2-3}
\end{equation}
The above characterizes a density operator if we also have $\|{\bf r}\| \leq 1$, as this condition is sufficient to ensure that the eigenvalues of $D$ are positive, hence that $D$ is positive. Also, if $\|{\bf r}\| = 1$, the density operator corresponds to a one-dimensional projection operator, i.e., it is representative of a pure state. This means that all vectors lying on the surface of the ball $B_1(\mathbb{R}^{3})$ are representative of pure states, and all vectors inside of it are genuine density operators, i.e., non-extremal convex combinations of one-dimensional projection operators. This bijection, between the density operators $D({\bf r})$ and the points of $B_1(\mathbb{R}^{3})$, is known as the Bloch geometrical representation of the state space of a two-dimensional quantum system (qubit), and is an expression of the well-known $SU(2)$-$SO(3)$ two-to-one homomorphism.

\section{Non-Born probabilities}
\label{generalized}

\noindent Gleason's theorem states that the Born rule can be derived in the standard quantum formalism by assuming that probabilities depend only on the projectors and are additive on mutually orthogonal projectors. However, Gleason's theorem does not hold for two-dimensional Hilbert spaces, and it is easy to show that, for  dimension $N=2$, infinitely many probability measures can be defined. 

The failure of Gleason's original theorem in dimension two should be distinguished from Gleason-type theorems obtained by strengthening the assumptions. Busch, for example, replaces projection-valued measures (PVM) by effects and obtains a theorem valid also for qubits (an effect is a positive operator $E$ satisfying $0 \le E \le \mathbb{I}$; in generalized measurements, or POVMs, measurement outcomes are represented by such operators rather than by projection operators); the stronger additivity requirement on effects then supplies constraints that are absent for PVMs  \cite{Busch2003}. Caves \emph{et al.} developed a related POVM formulation \cite{Caves2004}, while Wright and Weigert later showed that it is sufficient to consider projective measurements together with their classical mixtures \cite{WrightWeigert2019Qubit}. More recently, Fiorentino and Weigert derived a qubit extension that retains projective measurements but invokes the tensor-product structure of composite systems \cite{FiorentinoWeigert2026}.

De Zela proposed a different extension \cite{DeZela2016}, asserting that the reason Gleason's original proof fails in the two-dimensional case is that the additivity assumption is not sufficient to derive the continuity of the probability measure, and that discontinuous non-Born measures can therefore exist; see also \cite{Hall2016,DeZelaReply2016}. Although De Zela proposes a Gleason-type theorem for two-dimensional systems based on a theorem by Gudder \cite{Gudder1979}, the latter requires not only continuity but also the property of orthogonal additivity, which is significantly stronger than the additivity condition imposed on projection-valued measures. Indeed, as we now show, there exist infinitely many \emph{continuous} non-Born probability assignments satisfying the normalization conditions for every two-outcome projective measurement.

Let $\{|a_1\rangle, |a_2\rangle\}$ be the basis of eigenvector-states associated with an arbitrary observable $A=a_1 P_{a_1}+a_2 P_{a_2}$. According to (\ref{formula2X2-3}), the two projections $P_{a_1}=|a_1\rangle\langle a_1|$ and $P_{a_2}=|a_2\rangle\langle a_2|$  can be associated, respectively, to the unit vectors ${\bf n}_1$ and ${\bf n}_2$ in $B_1(\mathbb{R}^{3})$: 
\begin{equation}
P_{a_1} = {1\over 2}\left(\mathbb{I} +{\bf n}_1\cdot\mbox{\boldmath$\sigma$}\right),\quad P_{a_2} = {1\over 2}\left(\mathbb{I} +{\bf n}_2\cdot\mbox{\boldmath$\sigma$}\right)
\end{equation}
Taking into account (\ref{pauliproperties2}), the Born rule then tells us that the transition probability ${\cal P}_{\rm Born}({\bf r}\to  {\bf n}_i)$, from a density operator $D({\bf r})$ to an eigenstate $P_{a_i}$, is given by the trace:
\begin{eqnarray}
{\rm Tr}\left[D({\bf r})P_{a_i}\right]&=&{1\over 4}\left[{\rm Tr}\,\mathbb{I}+ {\rm Tr}({\bf r}\cdot\mbox{\boldmath$\sigma$}\,\,{\bf n}_i\cdot\mbox{\boldmath$\sigma$})\right]={1\over 4}\left[2+\sum_{j,k=1}^3 r_j[n_i]_k {\rm Tr}\,(\sigma_j\sigma_k)\right]  \nonumber \\
&=&{1\over 2} \left(1+ {\bf r}\cdot {\bf n}_i\right)
\label{Born}
\end{eqnarray}
Note that $P_{a_1}P_{a_2}=0$ implies ${\rm Tr} \, (P_{a_1}P_{a_2})=0$, and  from (\ref{Born}) we deduce that $ {\bf n}_1\cdot {\bf n}_2=-1$, i.e., that the two Bloch eigenvectors point in opposite directions in $B_1(\mathbb{R}^{3})$.

To define an illustrative family of generalized probability assignments, let $f:[-1,1]\to [-1,1]$ be any real-valued odd function, i.e., $f(-x)=-f(x)$, so that $f(0)=0$, and let us also assume that $f(1)=1$. Examples are $f(x) = x^{2n+1}$, $n\in\mathbb{N}$. Using functions of this kind (which may or may not be continuous), we can define the following non-Born transition probabilities: 
\begin{equation}
{\cal P}_f({\bf r}\to  {\bf n}_i)={1\over 2} \left[1+ f({\bf r}\cdot {\bf n}_i)\right]
\label{extended-Born}
\end{equation}
Indeed, from $f(1)=1$ we obtain reflexivity: ${\cal P}_f({\bf n}_i \to  {\bf n}_i)=1$, $i=1,2$, and $f$ odd guarantees that  ${\cal P}_f({\bf n}_1 \to  {\bf n}_2)={\cal P}_f({\bf n}_2 \to  {\bf n}_1)=0$. Since $f({\bf r}\cdot {\bf n}_1)=f(-{\bf r}\cdot {\bf n}_2)=-f({\bf r}\cdot {\bf n}_2)$, we also have additivity and normalization: ${\cal P}_f({\bf r}\to  {\bf n}_1)+{\cal P}_f({\bf r}\to  {\bf n}_2)=1$. In other words ${\cal P}_f$ defines a bona fide probability measure and the structure of a two-dimensional Hilbert space cannot force the Born rule, even if we impose continuity.

\section{Generalized Pauli matrices}
\label{generators}

\noindent In dimension $N\geq 3$, it is possible to generalize the previous construction. Instead of a three-dimensional unit Bloch sphere, completely filled with states, we now have a convex region of states contained in a unit ball $B_1(\real^{N^2 -1})$. To explain why the ball becomes $(N^2-1)$-dimensional, we can observe that a density operator is a $N\times N$ complex matrix. Being self-adjoint, the $N$ elements forming its diagonal are real numbers. These numbers must also sum up to $1$, so  we only need $N-1$ real parameters to specify the diagonal elements. Regarding the off-diagonal ones, the upper ones being the complex conjugate of the lower ones, we only have to determine ${1\over 2}(N^2-N)$ complex numbers, i.e., $N^2-N$ real numbers. Thus, the total number of real parameters needed to specify a general density operator $D$ is: $N^2-N + (N-1)= N^2-1$.

More precisely, we now need a set of $N^2$ matrices, $\{{\mathbb I},\Lambda_1,\dots,\Lambda_{N^2-1}\}$, forming an \emph{orthogonal basis} for  all linear operators acting on ${\mathbb C}^N$. We can ask them to be self-adjoint, traceless and orthogonal, i.e., to  generalize the properties of the Pauli matrices. More precisely, for all $i,j=1,\dots, N^2-1$, we require: 
\begin{equation}
\Lambda_i=\Lambda_i^\dagger,\quad {\rm Tr}\, \Lambda_i =0,\quad {\rm Tr}\, \Lambda_i\Lambda_j=2\delta_{ij}
\label{propertyLambda}
\end{equation}
\newline
Given an orthonormal basis $\{|b_1\rangle,\dots,|b_N\rangle\}$, a possible construction of these $N^2-1$ generalized Pauli matrices is the following~\cite{Hioe1981, Alicki1987, Mahler1995, AertsSassoli2014}:
\begin{eqnarray}
\label{rgeneratorsN}
&&\{\Lambda_i\}_{i=1}^{N^2-1}=\{U_{jk},V_{jk},W_{\ell}\},\quad 1\leq j < k\leq N,\quad 1\leq \ell\leq N-1\nonumber\\
&&U_{jk}=|b_j\rangle\langle b_k| + |b_k\rangle\langle b_j|, \quad V_{jk}=-i(|b_j\rangle\langle b_k| - |b_k\rangle\langle b_j|)\nonumber\\
&&W_\ell =-\sqrt{2\over \ell(\ell +1)}\left(\ell |b_{\ell +1}\rangle\langle b_{\ell +1}| - \sum_{j=1}^ \ell |b_j\rangle\langle b_j|\right)
\label{generatorsW}
\end{eqnarray}
Any density operator $D$ can then be written as the linear combination:
\begin{equation}
D({\bf r}) =  {1\over N}\left(\mathbb{I} +c_N\, {\bf r}\cdot\mbox{\boldmath$\Lambda$}\right)= {1\over N}\left[\mathbb{I} + \sqrt{\tfrac{N(N-1)}{2}}\sum_{i=1}^{N^2-1} r_i \Lambda_i\right]
\label{formulaNxN}
\end{equation}
where all the components $r_i$ of the $(N^2-1)$-dimensional vector ${\bf r}$ are real numbers, so that $D({\bf r})$ is manifestly self-adjoint, and considering that the $\Lambda_i$ obey (\ref{propertyLambda}), we also have 
\begin{equation}
{\rm Tr}\, D({\bf r}) =1,\quad \sqrt{N\over 2(N-1)}{\rm Tr}\left[D({\bf r})\Lambda_i\right]  =  r_i
\label{components-N}
\end{equation}
To guarantee that an operator written as in (\ref{formulaNxN}) is a density operator, it must also be positive semidefinite, but this will no longer be the case for any choice of a vector ${\bf r}\in B_1(\real^{N^2 -1})$. This is why only a convex portion of the unit ball is now filled with states; see \cite{Hioe1981, Alicki1987, Mahler1995, AertsSassoli2014} for more details.

\section{Forcing the Born rule}
\label{Transition}

\noindent Let us now derive the Born transition probabilities, generalizing formula (\ref{Born}) to the $N\geq 3$ case. Subsequently, we will try to generalize it by introducing a real-valued function $f$, as we did in (\ref{extended-Born}) for the $N=2$ case. Let $\{|a_1\rangle, \dots, |a_N\rangle\}$ be the basis of eigenvector-states associated with an arbitrary observable $A=\sum_{i=1}^N a_i P_{a_i}$,  $P_{a_i}=|a_i\rangle\langle a_i|$, $i=1,\dots, N$. To each of the $P_{a_i}$, we can associate a unit vector ${\bf n}_i \in B_1(\mathbb{R}^{N^2-1})$, such that 
\begin{equation}
P_{a_i} = {1\over N}\left(\mathbb{I} +c_N\, {\bf n}_i\cdot\mbox{\boldmath$\Lambda$}\right)
\end{equation}
Taking into account (\ref{propertyLambda}), the Born probabilities are given by:
\begin{eqnarray}
{\rm Tr}\left[D({\bf r})P_{a_i}\right]  &=& {1\over N^2}\left[ {\rm Tr}\, \mathbb{I}+c_N^2\,{\rm Tr}\left({\bf r}\cdot\mbox{\boldmath$\Lambda$}\,\, {\bf n}_i\cdot\mbox{\boldmath$\Lambda$} \right)\right] ={1\over N^2}\left[N + c_N^2\sum_{j,k=1}^{N^2-1} r_j[n_i]_k {\rm Tr}\left(\Lambda_j\Lambda_k\right) \right]\nonumber\\
&=& {1\over N} \left[1+ (N-1)\,{\bf r}\cdot {\bf n}_i\right]
\label{transitiongeneralNxN}
\end{eqnarray}
Since all the projections $\{P_{a_1},\dots,P_{a_N}\}$ are mutually orthogonal, if ${\bf r}={\bf n}_j$, we have $D({\bf r}) =P_{a_j}$, ${\rm Tr}(P_{a_j}P_{a_i})=\delta_{ji}$, and from (\ref{transitiongeneralNxN}) we find 
\begin{equation}
{\bf n}_i \cdot {\bf n}_j = -{1\over N-1}+\delta_{ij} {N\over N-1}
\label{scalar-n}
\end{equation}
This means that the angle between ${\bf n}_i$ and ${\bf n}_j$, $i\neq j$, is $\theta_N= \cos^{-1} (-{1\over N-1})$, and that the ${\bf n}_i$ form the $N$ vertex-vectors of a $(N-1)$-simplex $\triangle_{N-1}$, whose center coincides with the center of the unit ball $B_1(\mathbb{R}^{N^2-1})$ in which it is inscribed. 

A simpler expression for the Born transition probabilities (\ref{transitiongeneralNxN}) can be derived by writing ${\bf r}$ as the sum ${\bf r} = {\bf r}^\perp + {\bf r}^\parallel$, where ${\bf r}^\parallel$ is the vector obtained by orthogonally projecting ${\bf r}$ onto the $(N-1)$-dimensional subspace generated by $\triangle_{N-1}$. Since ${\bf r}^\perp\cdot {\bf n}_i =0$, (\ref{transitiongeneralNxN}) becomes:
\begin{equation}
 {\rm Tr}\left[D({\bf r})P_{a_i}\right]= {1\over N} \left[1+ (N-1)\,{\bf r}^\parallel\cdot {\bf n}_i\right]
\label{transitiongeneralNxN-bis}
\end{equation}
Writing ${\bf r}^\parallel$ as a combination of the vertex vectors
\begin{equation}
{\bf r}^\parallel =\sum_{i=1}^{N} r^\parallel_i \,{\bf n}_i,\quad \sum_{i=1}^{N} {\bf n}_i=0, \quad \sum_{i=1}^{N} r^\parallel_i=1, \quad r^\parallel_i \ge 0
\label{rparallelexpansion}
\end{equation}
then using (\ref{scalar-n}), we obtain
\begin{equation}
 {\bf r}^\parallel \cdot {\bf n}_i =\sum_{j=1}^{N} r^\parallel_j \,{\bf n}_j \cdot {\bf n}_i =\sum_{j=1}^{N} r^\parallel_j\left( -{1\over N-1}+\delta_{ij} {N\over N-1}\right)
\label{rparallelnj}
\end{equation}
Inserting (\ref{rparallelnj}) into (\ref{transitiongeneralNxN-bis}), then using (\ref{rparallelexpansion}), the Born transition probabilities become:
\begin{equation}
{\rm Tr}\left[D({\bf r})P_{a_i}\right]  =  r^\parallel_i
\label{trans-general}
\end{equation}

Let us now apply the same strategy as in the $N = 2 $ case, trying to generalize the Born rule (\ref{transitiongeneralNxN}) by introducing an auxiliary real function $f$. So, we tentatively introduce the generalized transition probabilities
\begin{equation}
{\cal P}_f({\bf r}\to{\bf n}_i)
=\frac{1}{N}\Big[1+(N-1)f({\bf r}\!\cdot\!{\bf n}_i)\Big]
\label{generalized-probability}
\end{equation}
The condition $0\le {\cal P}_f\le1$ implies that $f$ is a bounded function such that
\begin{equation}
-\frac{1}{N-1}\le f(x)\le1, \qquad x\in\Big[-\frac{1}{N-1},1\Big]
\end{equation}
Reflexivity imposes the condition $f(1)=1$, and additivity (plus normalization) requires $\sum_{i=1}^N f({\bf r}\!\cdot\!{\bf n}_i)=0$. Using (\ref{transitiongeneralNxN-bis}) and (\ref{trans-general}), this becomes the functional constraint
\begin{equation}
\sum_{i=1}^N
f\!\left(\frac{N r_i^\parallel-1}{N-1}\right)=0
\label{condition-1}
\end{equation}

We can therefore observe that the crucial new feature for $N\geq3$ is that the geometry of a projective measurement imposes stronger constraints on the allowed probability assignments than in the qubit case. For $N=2$, the two Bloch vectors associated with an orthonormal basis are antipodal, so normalization only requires a relation between $f(x)$ and $f(-x)$. In higher dimensions, the $N$ Bloch vectors form a regular simplex and the corresponding scalar products are linked by additional geometric relations, leading to a functional relation involving several arguments of $f$, which ultimately reduces, as we are going to see, to a Cauchy-type equation. This mechanism is of course related to the standard role of \emph{intertwining} in Gleason's theorem, where the same rank-one projector may belong to different orthogonal resolutions of the identity \cite{Gleason1957,Cooke1985,Dvurecenskij1996,WrightWeigert2019Cauchy}. In the present derivation, however, intertwining is not used explicitly and the corresponding restriction emerges directly from the geometry of the measurement simplex.

Choosing $r_i^\parallel=\frac1N$, $i=1,\dots,N$ (corresponding to a vector at the center of the simplex), the constraint (\ref{condition-1}) becomes $Nf(0)=0$, implying $f(0)=0$. Taking instead $r_1^\parallel=a$, $r_2^\parallel=b$, $r_3^\parallel=1-a-b$, $r_i^\parallel=0$ for $i\ge4$, with $a,b\ge0$ and $a+b\le1$ (corresponding to a vector belonging to a two-dimensional face of the simplex), (\ref{condition-1}) becomes
\begin{equation}
f\!\left(\frac{Na-1}{N-1}\right)
+f\!\left(\frac{Nb-1}{N-1}\right)
+f\!\left(\frac{N(1-a-b)-1}{N-1}\right)
+(N-3)f\!\left(-\frac1{N-1}\right)=0
\label{condition-3}
\end{equation}
Setting $a=1$ (and hence $b=0$) in (\ref{condition-3}), we find $f(-\frac1{N-1})=-\frac1{N-1}$. It is useful at this point to introduce the variables
\begin{equation}
x=\frac{Na-1}{N-1}, \quad y=\frac{Nb-1}{N-1}
\end{equation}
Eq.~(\ref{condition-3}) then becomes
\begin{equation}
f(x)+f(y)+f(\alpha-x-y)=\alpha,\qquad \alpha=\frac{N-3}{N-1}
\label{triangular-dominion}
\end{equation}
Note that   $a,b\in [0,1]$ is equivalent to $x,y\in [-{1\over N-1},1]$. Also,  $a+b\in [0,1]$ is equivalent to $x+y\in [-{2\over N-1},{N-2\over N-1}]$. Assume now for simplicity that $f$ is differentiable. Differentiating (\ref{triangular-dominion}) with respect to $x$ gives
\begin{equation}
f'(x)-f'(\alpha-x-y)=0
\end{equation}
Since $y$ can be varied, it immediately follows that $f'$ is a constant, hence $f(x) = cx + d$. From $f(0)=0$ we obtain $d=0$, and from  $f(1)=1$ we obtain $c=1$. Hence
\begin{equation}
f(x)=x
\label{affine}
\end{equation}
In other words, the generalized rule (\ref{generalized-probability}) reduces uniquely to the Born rule (\ref{transitiongeneralNxN}).

\section{Relaxing differentiability}

Differentiability of the function $f$ is not a necessary ingredient to obtain (\ref{affine}). Equation~(\ref{triangular-dominion}) holds on the triangular domain $x\geq -L$, $y\geq -L$, $x+y\leq 1-L$, where we have defined $L=\frac{1}{N-1}$. Setting $x=0$ in (\ref{triangular-dominion}) and using $f(0)=0$, we therefore obtain $f(y)=\alpha-f(\alpha-y)$, $-L\leq y\leq 1-L$. Using this relation in (\ref{triangular-dominion}) gives
\begin{equation}
f(x)+f(\alpha-x-y)=f(\alpha-y).
\label{transport}
\end{equation}
Introducing the variable $z=\alpha-x-y$, we have $\alpha-y=x+z$, and hence
\begin{equation}
f(x)+f(z)=f(x+z)
\label{local-cauchy}
\end{equation}
The allowed triangular domain implies $x\geq -L$, $z\geq -L$, $x+z\leq 1-L$, so that (\ref{local-cauchy}) is a local Cauchy functional equation on this domain.

Since $f$ is bounded, the standard regularity result for local solutions of the Cauchy equation applies \cite{WrightWeigert2019Cauchy,Aczel1966}. For completeness, let us first show continuity at $x=0$. Equation~(\ref{local-cauchy}) holds for $x$ and $z$ varying independently in an $\epsilon$-neighborhood of the origin. Indeed, since $z=\alpha-x-y$, for arbitrary $x,z\in[-\epsilon,\epsilon]$ we may choose $y=\alpha-x-z$; if $\epsilon>0$ is sufficiently small, this value of $y$ remains within the allowed domain, so that (\ref{transport}) applies and yields (\ref{local-cauchy}). Let then $x\in[-{\epsilon\over n},{\epsilon\over n}]$, for some positive integer $n$. Eq.~(\ref{local-cauchy})  gives $f(2x)=f(x+x)=f(x)+f(x)=2f(x)$, $f(3x)=f(x+2x)=f(x)+f(2x)=3f(x)$, and so on, and by induction we obtain $f(nx)=nf(x)$, since all the $kx$, $k=1,\dots,n$, belong to the interval $[-\epsilon,\epsilon]$, where (\ref{local-cauchy}) holds. This means that $|x|\leq{\epsilon\over n}$ implies
\begin{equation}
|f(x)|=\frac{|f(nx)|}{n}\le\frac1n
\end{equation}
showing continuity at $0$. Continuity throughout the interval $[-\epsilon,\epsilon]$ immediately follows. Indeed,
 \begin{equation}
\lim_{h\to 0}f(x+h)= f(x)+\lim_{h\to 0}f(h)=f(x)
\label{continuity}
\end{equation} 

The last step is to show that continuity on this interval implies linearity. We have $mf(x)=f(mx)=f(n{m\over n}x)=nf({m\over n}x)$, implying $f({m\over n}x)={m\over n}f(x)$, $m,n\in\mathbb{N}$, whenever the arguments involved remain in the interval where (\ref{local-cauchy}) holds. Also, $0=f(0)=f(x-x)=f(x)+f(-x)$, implying that $f(x)$ is an odd function; hence $f(rx)=rf(x)$, $r\in\mathbb{Q}$. By continuity and the density of the rational numbers in the real line, this relation extends to real values of $r$, showing that 
\begin{equation}
f(x)=cx,\quad x\in[-\epsilon,\epsilon]
\label{cx}
\end{equation}
for some $\epsilon>0$. Let us now show that this extends to the full physical domain $[-L,1]$ of the function $f$. For this, let $t\in[0,1-L]$, and choose a positive integer $n$ sufficiently large that $\frac{t}{n}\in[0,\epsilon]$. Since all the points $kt/n$, with $k=1,\ldots,n$, belong to $[0,1-L]$, repeated application of (\ref{local-cauchy}) gives
\begin{equation}
f(t)=f(n\tfrac{t}{n})=n f(\tfrac{t}{n})=n\,c\tfrac{t}{n}=c\,t,
\end{equation}
where for the penultimate equality we have used (\ref{cx}). We can reason in the same way if $t\in[-L,0]$, choosing $n$ sufficiently large that $t/n\in[-\epsilon,0]$. Hence, we have extended (\ref{cx}) to the interval $[-L,1-L]$.

It remains to extend this result to the interval $[1-L,1]$. We first determine the constant $c$. Since $-L\in[-L,1-L]$, the linearity already established on this interval gives $f(-L)=-cL$. On the other hand, we have previously shown that $f(-L)=-L$. Since $L>0$, it follows immediately that $c=1$. Thus,
\begin{equation}
f(x)=x,\quad x\in[-L,1-L]
\label{identity-first-domain}
\end{equation}
Now let $t\in[1-L,1]$. Setting $x=t$ and $y=-L$ in (\ref{triangular-dominion}), the third argument becomes $\alpha-t+L=1-L-t$, which belongs to $[-L,0]\subset [-L,1-L]$, i.e., to an already established interval. Eq.~(\ref{triangular-dominion}) then gives
\begin{equation}
f(t)+f(-L)+f(1-L-t)=1-2L
\end{equation}
Using (\ref{identity-first-domain}), we find
\begin{equation}
f(t)-L+1-L-t=1-2L
\end{equation}
It follows immediately that (\ref{identity-first-domain}) also holds in $[1-L,1]$. Combining the two intervals, we finally obtain
\begin{equation}
f(x)=x,\quad x\in[-L,1]
\end{equation}
which proves that the local linear solution extends to the full physical domain of $f$.

\section{Conclusion}

\noindent Gleason's theorem is often interpreted as showing that the Born rule follows uniquely from the structure of Hilbert space. The present analysis clarifies the geometric origin of this result using the Bloch representation of quantum states. For two-dimensional systems the state space is the three-dimensional unit Bloch ball and measurement outcomes correspond to antipodal points on its surface. In this situation, the only constraint is that the two probabilities sum to one. This leaves complete freedom in the functional dependence on the scalar products ${\bf r}\cdot{\bf n}_i$, allowing infinitely many non-Born probability rules, even if continuity is required.

In dimensions $N\ge3$, however, the geometry changes qualitatively. The eigenstates of a measurement correspond to the vertices of a regular $(N-1)$-simplex inscribed in the generalized $(N^2-1)$-dimensional unit Bloch ball. The requirement that probabilities associated with these $N$ mutually orthogonal projectors sum to unity then imposes strong functional constraints. These constraints force the probability rule to be linear in the scalar products ${\bf r}\cdot{\bf n}_i$, uniquely recovering the Born rule.

Within the restricted class of scalar-product deformations considered here, this provides a geometric account of the dimensional distinction that is central to Gleason's theorem. The argument is not a proof of the theorem in its full generality, since the probability assignments are assumed from the outset to depend on the state and the projector through the Bloch-space scalar product. Its value is instead to show, in an elementary setting, how the higher-dimensional simplex geometry turns the normalization requirement into a functional constraint strong enough to eliminate nonlinear alternatives to the Born rule. In this sense, the construction offers a complementary geometric perspective on the mechanism underlying Gleason's result.

\end{document}